\documentclass[12pt]{article}
\usepackage{amssymb}
\usepackage{graphicx}
\def\be{\begin{equation}}
\def\ee{\end{equation}}
\begin{document}
\begin{center}
\hfill \vbox{
\hbox{DFCAL-TH 02/1}
\hbox{March 2002}}
\vskip 0.5cm
{\Large\bf Finite-size scaling and deconfinement transition: 
the case of 4D SU(2) pure gauge theory}
\end{center}
\vskip 0.6cm
\centerline{Alessandro Papa and Carlo Vena}
\vskip 0.6cm
\centerline{\sl Dipartimento di Fisica, Universit\`a della Calabria}
\centerline{\sl \& Istituto Nazionale di Fisica Nucleare, Gruppo collegato
di Cosenza}
\centerline{\sl I--87036 Arcavacata di Rende, Cosenza, Italy}
\vskip0.2cm
\centerline{\sl E-mail: papa,vena @cs.infn.it}
\vskip 0.6cm
\begin{abstract}
A recently introduced method for determining the critical indices of the
deconfinement transition in gauge theories, already tested for the case of 
3D SU(3) pure gauge theory, is applied here to 4D SU(2) pure gauge theory.
The method is inspired by universality and based on the finite size scaling 
behavior of the expectation value of simple lattice operators, such as the 
plaquette. We obtain an accurate determination of the critical index $\nu$, 
in agreement with the prediction of the Svetitsky-Yaffe conjecture.
\end{abstract}

\newpage

\section{Introduction}
\label{sec:intro}
The phase transition from a low temperature phase, where quarks and gluons
are confined into hadrons, to a high temperature phase, where a quark-gluon plasma 
of interacting quasi-particles can appear, is one of the most important 
features of SU(N) gauge theories at finite temperature. Many present 
investigations (for a review, see Ref.~\cite{Karsch:2001cy}) are devoted 
to clarify the kind of phase transition as a function of 
the number of light quarks and of their masses and to determine the 
critical temperature. When the phase transition is {\em second order}, it
becomes essential to determine its critical indices, since
they rule the dependence of macroscopic observables on the temperature in the 
critical region. Moreover, they allow to identify the universality class of 
the transition and, consequently, the relevant degrees of freedom near criticality.

Pure SU(N) gauge theories at finite temperature possess all essential features 
of the deconfinement transition and represent an ideal laboratory to understand 
mechanisms and to set up methods of investigation, since they can be simulated on
a space-time lattice with relatively small effort.

The order parameter of the deconfinement transition in a SU(N) pure gauge
theory is the Polyakov loop~\cite{Polyakov:vu,Susskind:up}, 
defined as
\begin{equation}
P(\vec x) \equiv \mbox{Tr} \prod_{n_4=1}^{N_t} U_4(\vec x, a n_4)\;,
\label{Polyakov}
\end{equation}
where $U_\mu(x)$ is the link variable at the site $x\equiv(\vec x,t)$ in the $\mu$ 
direction and $a$ is the lattice spacing. Let us consider the transformation 
$U_4(\vec x, a \bar n_4) \longrightarrow z \: U_4(\vec x, a \bar n_4)$, where 
$z$ belongs to the {\em center} of the gauge group, Z(N), and 
$a \bar n_4$ is any fixed (lattice) time coordinate. Under this transformation
the lattice action is left invariant, the Polyakov loop $P(\vec x)$ transforms 
instead in $z \: P(\vec x)$. If the Z(N) symmetry were unbroken, 
$\langle P \rangle$ should always vanish; instead, owing to spontaneous 
symmetry breaking, $\langle P \rangle \neq 0$ above a critical temperature $T_c$.
The deconfinement transition coincides with the spontaneous breaking of the Z(N)
global symmetry.

By integrating out all degrees of freedom of the $(d+1)$-dimensional SU(N)
gauge theory except the order parameter, it is possible to generate 
an effective $d$-dimensional statistical model for the Polyakov loop, having Z(N)
as global symmetry group. According to the Svetitsky-Yaffe 
conjecture~\cite{Svetitsky:1982gs},
if the $(d+1)$-dimensional gauge theory undergoes a second order phase 
transition, its critical indices coincide with those of the effective 
$d$-dimensional model, if the latter has also a second order phase transition. 
Moreover their critical behavior, including finite size scaling, is predicted 
to coincide and is determined essentially by the Z(N) global symmetry. So, 
in particular, 3D SU(3) pure gauge theory belongs to the 
universality class of 2D Z(3) (3-state Potts) model and 4D SU(2) pure 
gauge theory to that of 3D Z(2) (Ising) model.

A new method has recently been proposed~\cite{Fiore:2001pf} 
(see also~\cite{Fiore:2001ci}) for the computation 
of the critical indices of a pure gauge theory with second order deconfinement 
transition. The method, described in detail in the next Section, is based on the 
finite size behavior of ``simple'' lattice operators, such as the plaquette, and 
allows very accurate determinations with small computational effort. It has 
been already successfully tested for the computation of the critical index 
$\nu$ of the correlation length for the case of 3D SU(3) pure gauge 
theory~\cite{Fiore:2001pf}. In that case a precise determination of $\nu$ 
was obtained, in perfect agreement with the Svetitsky-Yaffe
conjecture and with remarkably improved accuracy with respect to an earlier 
determination based on the Monte Carlo method~\cite{Engels:1997dz}. Here, we 
apply the same method to determine $\nu$ for the case of 4D SU(2) pure gauge theory.
On the basis of the previous experience~\cite{Fiore:2001pf}, we expect to 
improve the accuracy with respect to the existing 
determination~\cite{Fortunato:2000hg}, based on Monte Carlo methods, which gave 
$\nu_{MC}$=0.630(9).   

\section{Finite size scaling and the critical index $\nu$}\footnote{The 
content of this Section is not original and has been taken to a large extent from 
Ref.~\cite{Fiore:2001pf}. The reader who is already aware of that 
work can go directly to the next Section.}
\label{sec:nu}
To illustrate the method, let us consider the determination of the critical
index $\nu$ of the correlation length $\xi$, defined as
\[
\xi \sim |t|^{-\nu}\;, \hspace{3cm} t\equiv \frac{T-T_c}{T_c}\;.
\]
For a general $d$-dimensional statistical model a possible way of
extracting the value of $\nu$ from lattice Monte Carlo simulations is to study
the finite size scaling (FSS) behavior of the energy operator: 
FSS theory predicts that, if $L$ is the lattice size, the expectation value
of the (lattice) energy operator behaves for large $L$ as~\cite{Hasenbusch:1997}
\be
\langle E\rangle_L\sim \langle E\rangle_\infty+k L^{\frac{1}{\nu}-d}\;,
\label{efss}
\ee
where $\langle E\rangle_\infty$ is the  expectation value of the
energy operator in the thermodynamic limit and $k$ is a non-universal
constant.

Compared to other methods based on the FSS of fluctuation operators
such as susceptibilities or Binder cumulants, the advantage lies in
the fact that $\langle E\rangle_L$ can be computed to high
accuracy. The main drawback is that the term containing $\nu$ in
Eq.(\ref{efss}) is subdominant for $L\to\infty$ with respect to the
bulk contribution $\langle E\rangle_\infty$. The numerical results of
Ref.~\cite{Fiore:2001pf} have clearly shown that, in the case of gauge 
theories, the advantages outweigh the drawbacks and the method can give 
very accurate results.  

In order to apply the same method to the $(d+1)$-dimensional gauge
theory, it is necessary to compute expectation values of operators 
having the same finite size behavior of the energy in the $d$-dimensional
statistical model. Let us consider a gauge invariant operator $\hat{O}$
that is invariant also under the global symmetry given by the center of 
the gauge group, such as any Wilson loop or any correlator of the form
\be
P(\vec x) P^{\dagger}(\vec y)\;,
\label{pp}
\ee
where $P(\vec x)$ and $P(\vec y)$ are Polyakov loops at two different sites
$\vec x$ and $\vec y$ of the $d$-dimensional space. It is natural to expect 
that the operator product expansion (OPE) of any such operator at 
criticality has the same form as the one of $E$:
\be
\hat{O}=c_I\ I + c_\epsilon\  \epsilon 
+\cdots \;.
\label{ope}
\ee
Here $I$ and $\epsilon$ are, respectively, the identity and the {\em scaling} 
energy operator in the statistical model, and the dots represent contributions of 
operators with higher dimension. Two remarks are in order here. First, the 
scaling energy operator $E$ is not to be confused with the {\em lattice} energy 
operator $E$, which is just an example of an operator in the statistical model 
with OPE given by Eq.~(\ref{ope}). Second, the operators appearing in 
Eq.~(\ref{ope}) are, formally, functionals of the order parameter of the 
$(d+1)$-dimensional gauge theory, i.e. of the Polyakov loop. The dynamics of the 
order parameters at criticality is, however, the same in the $(d+1)$-dimensional 
gauge theory and in the $d$-dimensional statistical model, so that such a 
distinction is practically unessential.

The ansatz of Eq.~(\ref{ope}) was introduced and tested in 
Ref.~\cite{Gliozzi:1997yc}, and used in Ref.~\cite{Fiore:1998uk,Fiore:1998fs},
to obtain some exact results on correlation functions of 3D gauge theories 
at the deconfinement transition.
 
In particular, Eq.~(\ref{ope}) implies that the FSS behavior 
of the expectation value $\langle \hat{O} \rangle$ will have the form of 
Eq.~(\ref{efss}):
\be
\langle \hat{O}\rangle_L\sim \langle \hat{O}\rangle_\infty
+c L^{\frac{1}{\nu}-d}\;.
\label{ofss}
\ee
The contributions of the irrelevant operators will be subleading for 
$L\to \infty$.

We conclude that the FSS behavior of any such operator
can be used to determine the value of $\nu$ through
Eq.~(\ref{ofss}). The obvious advantage is that one can use operators,
such as the plaquette, whose expectation value can be computed to high
accuracy with relatively modest computational effort. 

In the next Section, we use the described method in practice, to determine the
critical index $\nu$ for the 4D SU(2) pure gauge theory.

\section{Numerical results}
\label{sec:res}
The lattice operators we considered are (i) time-like (``electric'') plaquette, 
(ii) space-like (``magnetic'') plaquette, (iii) the correlator 
$P(\vec x) P^{\dagger}(\vec y)$, with $\vec x$ and $\vec y$ taken to be nearest 
neighbors sites in the 3D (spatial) lattice\footnote{In SU(2) the Polyakov loop is 
a real quantity. The dagger in the definition of the Polyakov loop correlator 
has been kept for the sake of generality.}.
All these observables possess the required symmetry properties to have
an OPE given by Eq.~(\ref{ope}), and hence a FSS behavior as in Eq.~(\ref{ofss}). 
Moreover they can be computed to high accuracy in Monte Carlo simulations.

Expectation values were computed on lattices with temporal extension 
$N_t=2$ and spatial sizes $L=N_x=N_y=N_z$ ranging from $5$ to $26$. Simulations
must be performed at criticality, i.e. at the critical value of the lattice
(inverse) coupling constant $\beta$ for the given $N_t$. We fixed $\beta$ at the 
central value of the determination of Ref.~\cite{Fortunato:2000hg}: 
$\beta_c(N_t=2)$=1.8735. The simulation algorithm we adopted was the over-relaxed 
heat-bath~\cite{Petronzio:gp}.
For each simulation we collected a number of equilibrium configurations ranging
from 50K to 4.8M, according to the lattice size, separated each other by 
a number of iterations of the order of the autocorrelation time. The error 
analysis was performed by the jackknife method applied to data 
bins at different levels of blocking.

We report in Table~1 the expectation values we obtained. Because of the
asymmetry of the lattice, space-like and time-like plaquettes have
obviously different expectation values and must be considered as two
different operators.
To evaluate $\nu$, we performed a single multibranched fit of the
three data sets. To avoid cross-correlations we included
in the fit only magnetic plaquettes from lattices with odd $L$ and electric 
plaquettes from lattices with even $L$. Polyakov loop correlations were measured 
in separate runs and therefore are not correlated with the plaquette measurements.

The result of the fit is
\be
\nu=0.6298(28)\;, \qquad\qquad\chi^2/{\rm d.o.f.}=0.96\;,
\ee
in excellent agreement, as predicted by the Svetitsky-Yaffe conjecture,
with the most accurate determination for the 3D Ising model, 
$\nu=0.63012(16)$, obtained in Ref.~\cite{Campostrini:2002cf} 
by the high-temperature expansion method. According to expectations,
the accuracy of our result is improved in comparison to the existing Monte 
Carlo evaluation, based on the $\chi^2$ analysis of the topological loop 
susceptibility, which gives $\nu_{MC}=0.630(9)$~\cite{Fortunato:2000hg}. 

In Figs.~1-3, we plot the expectation values of electric plaquette, magnetic
plaquette and Polyakov loop correlator for different values of $L$ versus 
$L^{1/\nu-3}$, where $\nu$ is (the central value of) our determination.
 
\section{Conclusions}
\label{sec:concl}
In this paper, we have applied a recently proposed method for the 
evaluation of critical indices of the deconfinement transition in pure
gauge theories to the case of 4D SU(2). Our result for
the critical index $\nu$ of the correlation length confirms the conclusion
of the analogous determination for the case of 3D SU(3) by
the same method. Namely, the proposed method allows very precise results for 
critical indices, since it is based on the evaluation of ``simple'' expectation
values, such as that of the plaquette.

The same method can be used to evaluate the critical coupling, which here 
was taken from the literature: one should perform simulations at different values 
of $\beta$ in the critical region and, for each of them, determine $\nu$ by the
fit with the finite size scaling law; the critical value of $\beta$ can be obtained
by comparing the $\chi^2$'s of the different fits.

\section*{Acknowledgments}
We are grateful to R.~Fiore and P.~Provero for many fruitful discussions.

\newpage

\begin{table}[ht]
\caption{Expectation values of electric plaquette, magnetic plaquette and
Polyakov loop correlator. Simulations were performed at $\beta=\beta_c$=1.8735 
on $L^3\times 2$ lattices.}
\label{mcrplaq2}
\begin{center}
\begin{tabular}{|c|c|c|c|c|}
\hline
$L$ & Statistics & Electric plaq. & Magnetic plaq. & $PP$ correlator \\
\hline
 5  & 4800K      & 0.477152(11)   & 0.4658468(88) & - \\
 6  & 2700K      & 0.475211(11)   & 0.4652584(87) & - \\
 7  & 1700K      & 0.473926(12)   & 0.4649083(86) & - \\
 7  & 1100K      &      -         &      -        & 0.40117(14) \\
 8  & 1200K      & 0.473002(12)   & 0.4646322(88) & - \\
 9  &  800K      & 0.472354(13)   & 0.4644188(99) & - \\
10  &  600K      & 0.471832(13)   & 0.4642710(82) & 0.37296(13) \\
11  &  450K      & 0.471449(13)   & 0.4641638(85) & - \\
12  &  350K      & 0.471127(13)   & 0.4640621(90) & - \\
13  &  280K      & 0.470850(13)   & 0.4639818(80) & 0.35969(14) \\
14  &  280K      & 0.470659(11)   & 0.4639314(80) & - \\
15  &  200K      & 0.470482(12)   & 0.4638766(78) & - \\
16  &  150K      & 0.470290(14)   & 0.4638078(86) & 0.35208(14) \\
17  &  120K      & 0.470169(12)   & 0.4637843(89) & - \\
18  &  100K      & 0.470065(11)   & 0.4637625(84) & - \\
19  &   90K      & 0.469948(13)   & 0.463722(10)  & 0.34735(15) \\
20  &   75K      & 0.469872(15)   & 0.4636941(75) & - \\
21  &   65K      & 0.469771(15)   & 0.4636668(84) & - \\
22  &   53K      & 0.469714(16)   & 0.4636391(97) & 0.34429(23) \\
23  &   50K      & 0.469626(14)   & 0.4636321(77) & - \\
24  &   50K      & 0.469597(15)   & 0.4636137(90) & - \\
25  &   50K      & -              & -             & 0.34165(16) \\
26  &   50K      & 0.469487(15)   & 0.463571(12)  & - \\
\hline
\end{tabular}
\end{center}
\end{table}

\newpage

\begin{figure}[tb]
\includegraphics[width=1.\textwidth]{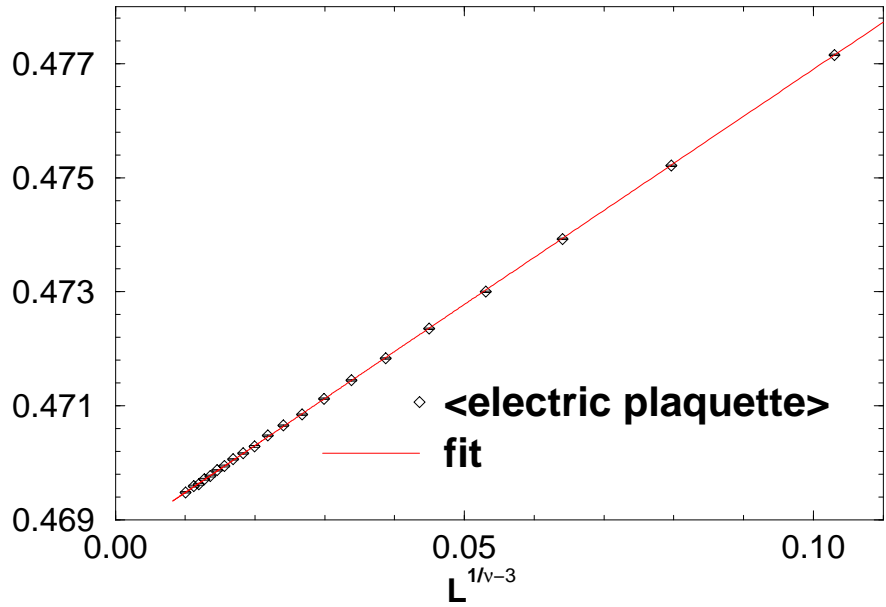}
\caption{Expectation values of the electric plaquette {\it vs} $L^{1/\nu-3}$, 
with $\nu$ taken as the central value of our determination, i.e. $\nu$=0.6298. 
Simulations were performed at $\beta=\beta_c$=1.8735 on $L^3\times2$ lattices,
with $L$ ranging from 5 to 26 (see Table~1).}
\end{figure}

\begin{figure}[bt]
\includegraphics[width=1.\textwidth]{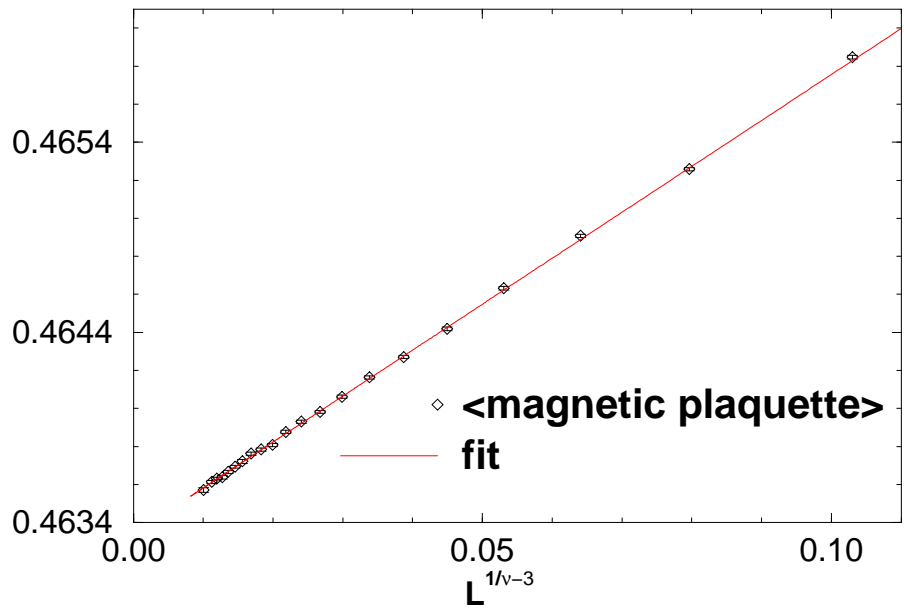}
\caption{The same as Fig.~1 for the magnetic plaquette.}
\end{figure}

\begin{figure}[tb]
\includegraphics[width=1.\textwidth]{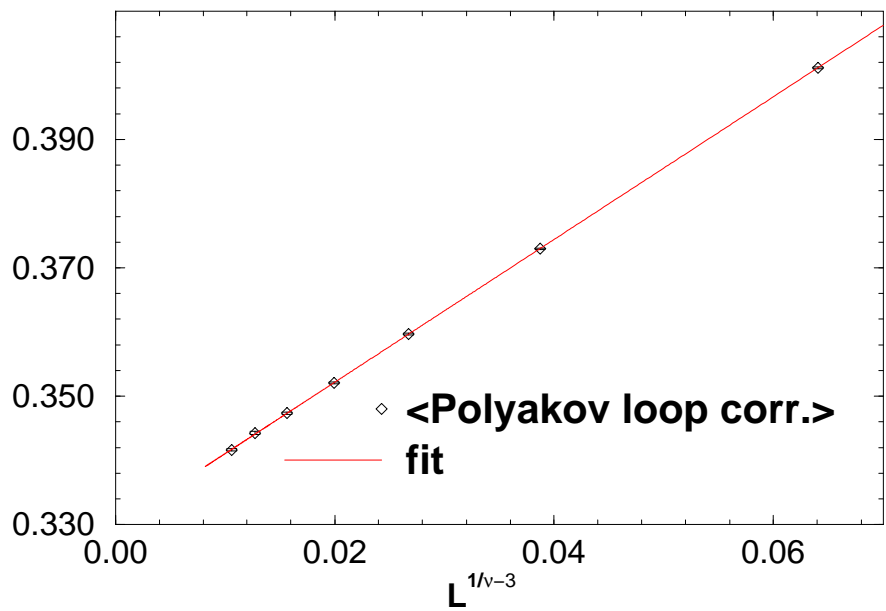}
\caption{The same as Fig.~1 for the Polyakov loop correlator.}
\end{figure}


\begin{thebibliography}{999}

\bibitem{Karsch:2001cy}
F.~Karsch,
arXiv:hep-lat/0106019.

\bibitem{Polyakov:vu}
A.~M.~Polyakov,
Phys.\ Lett.\ B {\bf 72} (1978) 477.

\bibitem{Susskind:up}
L.~Susskind,
Phys.\ Rev.\ D {\bf 20} (1979) 2610.

\bibitem{Svetitsky:1982gs}
B.~Svetitsky and L.~G.~Yaffe,
Nucl.\ Phys.\ B {\bf 210} (1982) 423.

\bibitem{Fiore:2001pf}
R.~Fiore, A.~Papa and P.~Provero,
Phys.\ Rev.\ D {\bf 63} (2001) 117503
[arXiv:hep-lat/0102004].

\bibitem{Fiore:2001ci}
R.~Fiore, A.~Papa and P.~Provero,
Nucl.\ Phys.\ Proc.\ Suppl.\  {\bf 106} (2002) 486
[arXiv:hep-lat/0110017].

\bibitem{Engels:1997dz}
J.~Engels, F.~Karsch, E.~Laermann, C.~Legeland, M.~Lutgemeier, B.~Petersson and 
T.~Scheideler,
Nucl.\ Phys.\ Proc.\ Suppl.\ {\bf 53} (1997) 420
[arXiv:hep-lat/9608099].

\bibitem{Fortunato:2000hg}
S.~Fortunato, F.~Karsch, P.~Petreczky and H.~Satz,
Nucl.\ Phys.\ Proc.\ Suppl.\  {\bf 94} (2001) 398
[arXiv:hep-lat/0010026].

\bibitem{Hasenbusch:1997}
M.~Hasenbusch and K.~Pinn, 
J.\ Phys.\ A {\bf 31} (1998) 6157 
[arXiv:cond-mat/9706003].

\bibitem{Gliozzi:1997yc}
F.~Gliozzi and P.~Provero,
Phys.\ Rev.\ D {\bf 56} (1997) 1131
[arXiv:hep-lat/9701014].

\bibitem{Fiore:1998uk}
R.~Fiore, F.~Gliozzi and P.~Provero,
Phys.\ Rev.\ D {\bf 58} (1998) 114502
[arXiv:hep-lat/9806017].

\bibitem{Fiore:1998fs}
R.~Fiore, F.~Gliozzi and P.~Provero,
Nucl.\ Phys.\ Proc.\ Suppl.\  {\bf 73} (1999) 429
[arXiv:hep-lat/9807037].

\bibitem{Petronzio:gp}
R.~Petronzio and E.~Vicari,
Phys.\ Lett.\ B {\bf 254} (1991) 444.

\bibitem{Campostrini:2002cf}
M.~Campostrini, A.~Pelissetto, P.~Rossi and E.~Vicari,
arXiv:cond-mat/0201180.

\end{thebibliography}
\end{document}